\documentclass[a4paper]{jpconf}
\usepackage[breaklinks,colorlinks,bookmarks=false,citecolor=blue,linkcolor=red,urlcolor=blue]{hyperref}
\usepackage{graphicx}
\usepackage{cite}
\usepackage{amsmath,amssymb}

\usepackage{color}
\definecolor{Blue}{rgb}{0.3,0.3,0.9}
\definecolor{Red}{rgb}{0.9,0.3,0.3}
\definecolor{Green}{rgb}{0.3,0.6,0.3}
\newcommand{\revision}[1]{{{#1}}}

\begin{document}
\title{Machine learning the 2D percolation model}

\author{Djénabou Bayo$^{1,2}$, Andreas Honecker$^{2}$, Rudolf A.\ R\"omer$^{1}$}

\address{$^{1}$Department of Physics, University of Warwick, Coventry, CV4 7AL,
United Kingdom}
\address{$^{2}$Laboratoire de Physique Th\'eorique et Mod\'elisation, CNRS UMR 8089, CY Cergy Paris Universit\'e, Cergy-Pontoise,  France}

\ead{djenabou.bayo@warwick.ac.uk}
%\ead{andreas.honecker@cyu.fr}
%\ead{r.roemer@warwick.ac.uk}

\begin{abstract}
We use deep-learning strategies to study the 2D percolation model on a square lattice. We employ standard image recognition tools with a multi-layered convolutional neural network. We test how well these strategies can characterise densities and correlation lengths of percolation states and whether the essential role of the percolating cluster is recognised.
\end{abstract}

%%%%%%%%%%%%%%%%%%%%%%%%%%%%%%%%%%%%%%%%%%%%%%%%%%%%%%%%%%%%%%%%%%%%%%%%%%%%%%%%%
\section{Introduction}
\label{sec:intro}
%%%%%%%%%%%%%%%%%%%%%%%%%%%%%%%%%%%%%%%%%%%%%%%%%%%%%%%%%%%%%%%%%%%%%%%%%%%%%%%%%

% A popular application of machine learning (ML) is the identification and classification of phases of matter \cite{Venderley2018MachineLO,Carrasquilla2017,PhysRevX.7.031038}. For example, many properties of the ferromagnetic-to-paramagnetic phase transition in the 2D Ising model were reconstructed in Ref.\ \cite{Carrasquilla2017} while Anderson-type transitions and phase diagrams for models of topological matter have been considered in Ref.\ \cite{doi:10.7566/JPSJ.89.022001}. 
%talk about S Acevedo, unsupervised
%
A popular application of machine learning (ML) is the identification and classification of phases of matter \cite{Venderley2018MachineLO,Carrasquilla2017,PhysRevX.7.031038,PhysRevE.99.032142}. For example, many properties of the ferromagnetic-to-paramagnetic phase transition in the 2D Ising model were reconstructed in Ref.\ \cite{Carrasquilla2017} while Anderson-type transitions and phase diagrams for models of topological matter have been considered in Ref.\ \cite{doi:10.7566/JPSJ.89.022001}. 
%
% \revision{The percolation model is one of the simplest model displaying a second order phase transition \cite{Stauffer2018}. It is a good candidate to study the performance of ML tools when looking at system displaying phase transitions. Previous studies  \cite{PhysRevE.99.032142,cheng2021machine} showed that supervised learning appears to identify well parameters of the percolation model. However, unsupervised learning and in particular PCA method fails to find the percolation transition. Furthermore, it also fails to obtain a correlation between the principal component and the order parameter \cite{cheng2021machine}, like it was shown to happen for the Ising model\cite{2017}. }
These promising results seem to pave the way for ML as a standard tool in condensed matter. 
%
%However while it was shown that ML process and methods like Convolutional Neural Network, usually used for image recognition allow to successfully classify phases, one might wonder about the real efficiency of those process when it comes to identifying transitions stemming from non local phenomena.
ML approaches based on deep, i.e., multi-layered, neural nets fall into three broad classes namely, \emph{supervised}, \emph{un-supervised} and \emph{reinforcement} learning approaches. Common to all approaches is the feed-forward/back-propagation cycle in which parameters of the neural nets are optimised such that suitably chosen loss functions become minimal. Various network architectures have been shown to be appropriate for tasks as varied as image recognition \cite{FUJIYOSHI2019244}, deepfakes \cite{1282,9156570} and board games \cite{Silver2017}. 
Here, we shall employ supervised learning strategies in the 2D percolation model on a square lattice. It is well known that the percolation in 2D exhibits a phase transition as a function of occupation densities with a diverging two-point correlation length at a critical density $p_c$.
\revision{%
%The percolation model is one of the simplest model displaying a second order phase transition \cite{Stauffer2018}. It is a good candidate to study the performance of ML tools when looking at system displaying phase transitions. 
Previous studies  \cite{PhysRevE.99.032142,cheng2021machine} found supervised learning able to distinguish the two phases of the percolation model, while the utility of unsupervised learning seems less established \cite{cheng2021machine,YU2020125065}. 
}%However, unsupervised learning and in particular PCA method fails to find the percolation transition. Furthermore, it also fails to obtain a correlation between the principal component and the order parameter \cite{cheng2021machine}, like it was shown to happen for the Ising model\cite{2017}. }
The aim of this paper is to develop a robust ML strategy to identify for finite lattices whether a given percolation lattice contains a spanning cluster or not. As such, we will (i) classify, with ML, percolation lattices according to densities and correlation lengths and (ii) use ML regression to predict densities and correlation lengths. 

%%%%%%%%%%%%%%%%%%%%%%%%%%%%%%%%%%%%%%%%%%%%%%%%%%%%%%%%%%%%%%%%%%%%%%%%%%%%%%%%%
\section{Model and Methods}
\label{sec:modelmethods}
%%%%%%%%%%%%%%%%%%%%%%%%%%%%%%%%%%%%%%%%%%%%%%%%%%%%%%%%%%%%%%%%%%%%%%%%%%%%%%%%%

%%%%%%%%%%%%%%%%%%%%%%%%%%%%%%%%%%%%%%%%%%%%%%%%%%%%%%%%%%%%%%%%%%%%%%%%%%%%%%%%%
%\subsection{Site percolation in small square lattices}

\begin{figure*}[ptb]
    \centering
     \raisebox{0.21\columnwidth}{(a)} \raisebox{0.04\columnwidth}{\includegraphics[width=0.19\columnwidth]{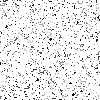}}
    \raisebox{0.21\columnwidth}{(b)} \raisebox{0.04\columnwidth}{\includegraphics[width=0.19\columnwidth]{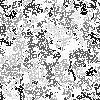}}
    \raisebox{0.21\columnwidth}{(c)} \raisebox{0.04\columnwidth}{\includegraphics[width=0.19\columnwidth]{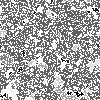}}
    \raisebox{0.21\columnwidth}{(d)}\hspace*{-2ex}%\vspace*{-2ex}
    \includegraphics[width=0.235\columnwidth]{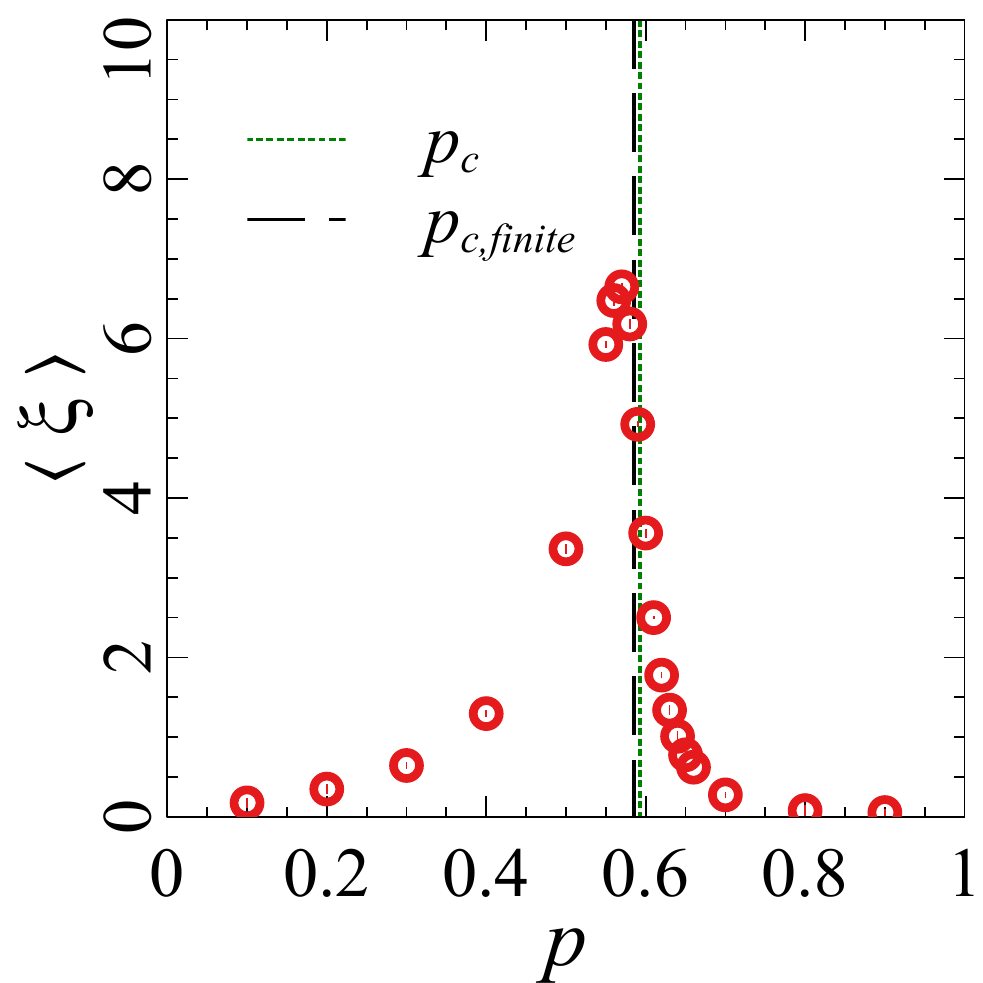}
\vspace*{-2.5mm}
\caption{
%%%% These are actual configurations, not schematic ones, aren't they?
%Schematic representation of
Percolation states on a $100 \times 100$ lattice with open boundary conditions
for (a) the non-percolating phase ($p=0.2 <p_c$),  
(b) close to the transition ($p=0.5 \approx p_c$) and (c) in the spanning phase ($p=0.6>p_c$). Different clusters of occupied sites are denoted by grey squares and identical grey levels indicate a common cluster. There are 1235, 597 and 248 clusters in (a), (b) and (c), resp., and the largest cluster size is 12, 252 and 4422, resp., in each panel.
(d) Average correlation length $\langle \xi \rangle(p)$ computed from $5000$ percolation states for $20$ $p$ values and \revision{periodic} boundaries. The vertical lines denote $p_c$ in the thermodynamic limit \cite{article} (green dotted) and the position $p=0.5853$ for the largest $\langle \xi \rangle$ as a estimate of $p_c$ for our finite-sized systems (black dashed). The standard error of the mean for each data point is within symbol size.}
    \label{fig:schematics}
\vspace*{-2.5mm}
\end{figure*}

The site percolation model on a 2D $L \times L$ lattice is defined as follows \cite{Stauffer2018}: \emph{sites} on the lattice are randomly occupied with probability $p$ and left empty with probability $1-p$. Each site has four \emph{neighbours}, i.e.,  two along the horizontal and two along the vertical axes. A group of neighbouring occupied sites is called a \emph{cluster}. For small $p$ values, most clusters have size $< L$ while for $p \rightarrow 1$, the cluster size tends to $L^2$. 
\revision{At the \emph{percolation threshold} $p_{c}$ a spanning cluster emerges which connects opposite sides of the lattice. Much previous work \cite{PhysRevLett.85.4104,Stauffer2018} has established that for $L \rightarrow\infty$, $p_{c}=0.59274605079210(2)$ \cite{article}.}
For finite $L$, one usually finds upon averaging a threshold value $p_c(L) < p_c$.
%since large but finite clusters with size $\gg L$ can be mistakenly identified as the percolation cluster. 
%
Numerically, the Hoshen-Kopelman cluster identification algorithm \cite{PhysRevB.14.3438} allows for a determination of $p_c(L)$ while also giving full information about the cluster geometries and shapes. 
We use it to compute configurations $\psi_n(\mathbf{r})$ of occupied and empty sites on $100 \times 100$ lattices with $20$ densities $p= 0.1, 0.2, \ldots, 0.55, 0.56, \ldots, 0.66, 0.7,\ldots, 0.9$ and $N=5000$ samples for each density. Here, $n$ labels different configurations while $\mathbf{r}$ %=(x,y)$ 
indicates %$x$ and $y$ 
positions of individual sites and $\psi_n(\mathbf{r})\in \{0,1\}$ denotes whether at $\mathbf{r}$ there is an empty ($0$) or occupied ($1$) site in the $n$ state. Examples of such $\psi_n$ are given in Fig.\ \ref{fig:schematics}(a-c). Our results suggest that ${p_c}(100)\approx 0.59
 \pm 0.03$ for our data, computed with open boundary conditions. 
The transition from non-spanning clusters at $p<p_c$ to a spanning cluster at $p > p_c$ is a continuous phase transition with the probability $P(p)$ that an arbitrary site belongs to the infinite cluster
%, with $P \propto (p-p_c)^{5/36}$, 
acting as the order parameter. Similarly, the \revision{connected} correlation function $g(r)$, defined as the probability of a site at distance $r$ from an occupied site to belong to the same cluster \cite{zhang2021machine}, diverges at $p_c$. Its associated correlation length
$%\begin{equation}\label{corr_l}
    \xi = \sqrt{{\sum_{r}r^2 g(r)}/{\sum_{r}g(r)}}
$ %\end{equation}
gives the average distance between two sites that belong to the same cluster.
As $p\rightarrow p_c$,  
$%  \begin{equation}
      \xi \propto \mid p-p_c \mid  ^{-4/3}
$ %  \end{equation}
%with $\nu=4/3$ the critical exponent 
for $L \rightarrow \infty$. An example of the behaviour of \revision{$\xi(r)$} in our data is shown in Fig.\ \ref{fig:schematics}(d) where we can see that $\xi$ indeed is maximal at ${p_c}(100)$ but does not diverge for the calculated $p$ values.
%%%%%%%%%%%%%%%%%%%%%%%%%%%%%%%%%%%%%%%%%%%%%%%%%%%%%%%%%%%%%%%%%%%%%%%%%%%%%%%%%

% \begin{figure*}[bt]
%     \centering
%     (a)\includegraphics[width=0.2\columnwidth]{Data info/correlation_l.png}
%     \caption{Average correlation length obtained with our dataset 
%     }
%     \label{fig:schematics}
% \end{figure*}
%%%%%%%%%%%%%%%%%%%%%%%%%%%%%%%%%%%%%%%%%%%%%%%%%%%%%%%%%%%%%%%%%%%%%%%%%%%%%%%%%

%%%%%%%%%%%%%%%%%%%%%%%%%%%%%%%%%%%%%%%%%%%%%%%%%%%%%%%%%%%%%%%%%%%%%%%%%%%%%%%%%
%\subsection{Machine learning}

Convolution neural nets (CNN) are a class of multi-layered (deep) neural nets in which spatial locality of data values is retained during the ML training. When coupled with a form of residual learning \cite{He2016DeepRL}, the resulting residual networks ({\sc ResNets}) have been shown to allow astonishing precision when classifying images, e.g., of animals \cite{https://doi.org/10.1111/2041-210X.13120} and handwritten characters \cite{Zhang2017CombinationOR}, or when predicting numerical values, e.g., of market prices \cite{Zhao2020WaveletDC}.
 \revision{Here, we shall use a {\sc ResNet18} \cite{He2016DeepRL} network with $17$ convolutional and $1$ fully-connected layers, pretrained on the {\sc ImageNet} dataset \cite{imagenet_cvpr09}. For our implementation we use the {\sc PyTorch} suite of ML codes \cite{NEURIPS2019_9015}.}
%- train/valid/test split
We train the {\sc ResNet18} on the $\psi_n$ configurations, using a $90\%$/$10\%$ split into training and validation data. Additional $\psi_n$  configurations are generated as necessary for test runs. 
%
%- loss
We concentrate on two ML tasks. First, we classify percolation configurations $\psi_n$ according to densities $p$, correlation lengths $\xi$ and spanning non-spanning. In the second task, we aim to predict $p$ and $\xi$ values via ML regression. In both problems, the overall network architecture remains identical, we just adapt the last layer. For the classification the output layers have a number of neurons corresponding to the number of classes trained i.e, for the classification by density the ${\cal C}=20$ values $p=0.1,0.2,\ldots,0.55,0.56,\ldots,0.66,0.7,0.8,0.9$, while for regression the output layer has only one neuron making the predictions.
However, the loss functions are different. 
Let $\omega$ denote the parameters (weights) of the {\sc ResNet} and let $(\psi_n,\chi_n)$ represent a given data sample with $\chi_n$ the classification/regression targets, i.e., classes $p$ or $\xi$, and also ${\chi}'_n$ the predicted values, $p'$ or $\xi'$. \revision{For classification of categorical data, the class names are denoted by a class index  $a= 1, \ldots, {\cal C}$ and encoded as
% $
% \chi_{ai}=\left\{
% \begin{array}{ll}
%       1 & \mbox{if }  \chi_{a}=i\\
%       0 & \mbox{otherwise}
%     \end{array}
%     \right.
%     $.
%
$\chi_{ai}=  1$  if $\chi_{ai}=i$, $0$ otherwise.}
Then, for the (multi-class) classification problem, we choose the usual 
cross-entropy loss function, $\l_\mathrm{c}(\omega)=-\sum_{i=1}^{N{\cal C}}  \sum_{a=1}^{{\cal C}} \chi_{ai} \log{\chi'}_{ai}(\omega)+(1-\chi_{ai})\log[1-\chi_{ai}(\omega)]$ \cite{Mehta2019AHL}.
%Here, e.g., when classifying for densities $p$, we can have at most ${\cal C}=20$ classes.
%
On the other hand, the loss function for the regression problem is given by the mean-squared error $l_\mathrm{r}(w)=\frac{1}{n} \sum_{i=1}^{N {\cal C}} [\chi_{i}- {\chi}'_i (w)]^2$.
When giving results for the values of the loss functions below, we always present those after averaging over at least $10$ independent training and validation cycles. 
\revision{We also represent the quality of a prediction by confusion matrices \cite{Mehta2019AHL}. These graphically represent the predicted class labels as a function of the true ones in matrix form, with an error-free prediction corresponding to a diagonal matrix.}
For comparison of the classification and regression calculations, we use in both cases a maximum number of $\epsilon_\mathrm{max}=20$ epochs. Our ML calculations train with a batch size of $16$ for classification and for regression on an NVIDIA Quadro RTX 6000 GPU card.

\section{Results} 
\label{sec:results}
%%%%%%%%%%%%%%%%%%%%%%%%%%%%%%%%%%%%%%%%%%%%%%%%%%%%%%%%%%%%%%%%%%%%%%%%%%%%%%%%%

%%%%%%%%%%%%%%%%%%%%%%%%%%%%%%%%%%%%%%%%%%%%%%%%%%%%%%%%%%%%%%%%%%%%%%%%%%%%%%%%%
%\subsection{Classification}

%%%%%%%%%%%%%%%%%%%%%%%%%%%%%%%%%%%%%%%%%%%%%%%%%%%%%%%%%%%%%%%%%%%%%%%%%%%%%%%%%
%\subsubsection{Density}

\begin{figure*}[tb]
    \centering%
    \raisebox{0.30\columnwidth}{(a)}\hspace*{+1ex}
    \includegraphics[width=0.48\columnwidth]{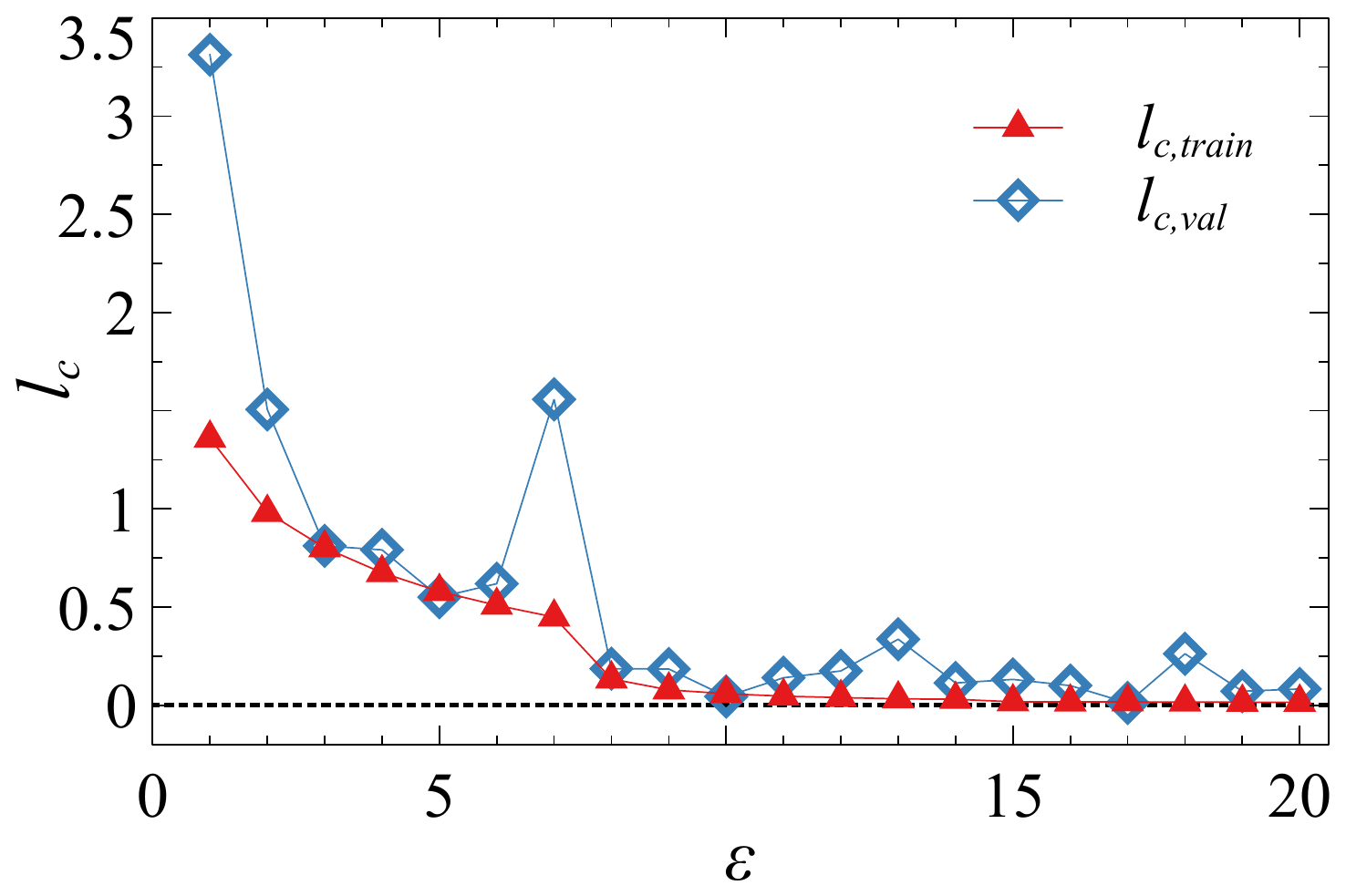}\
    \hspace{+3ex}
    \raisebox{0.30\columnwidth}{(b)}\hspace*{-1ex}
    \includegraphics[width=0.35\columnwidth]{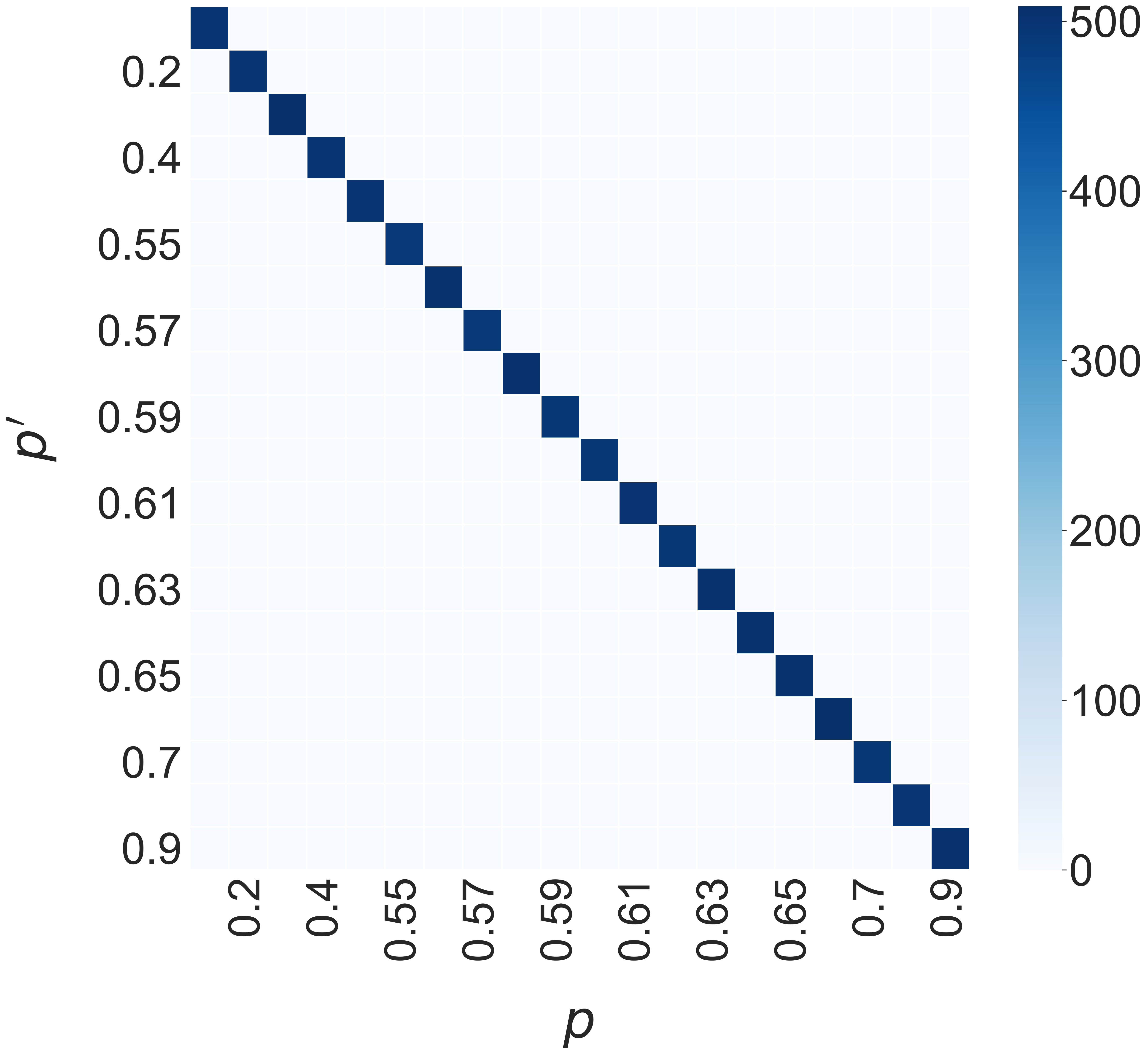}\\
    \raisebox{0.30\columnwidth}{(c)}\hspace*{+1ex}
    \includegraphics[width=0.48\columnwidth]{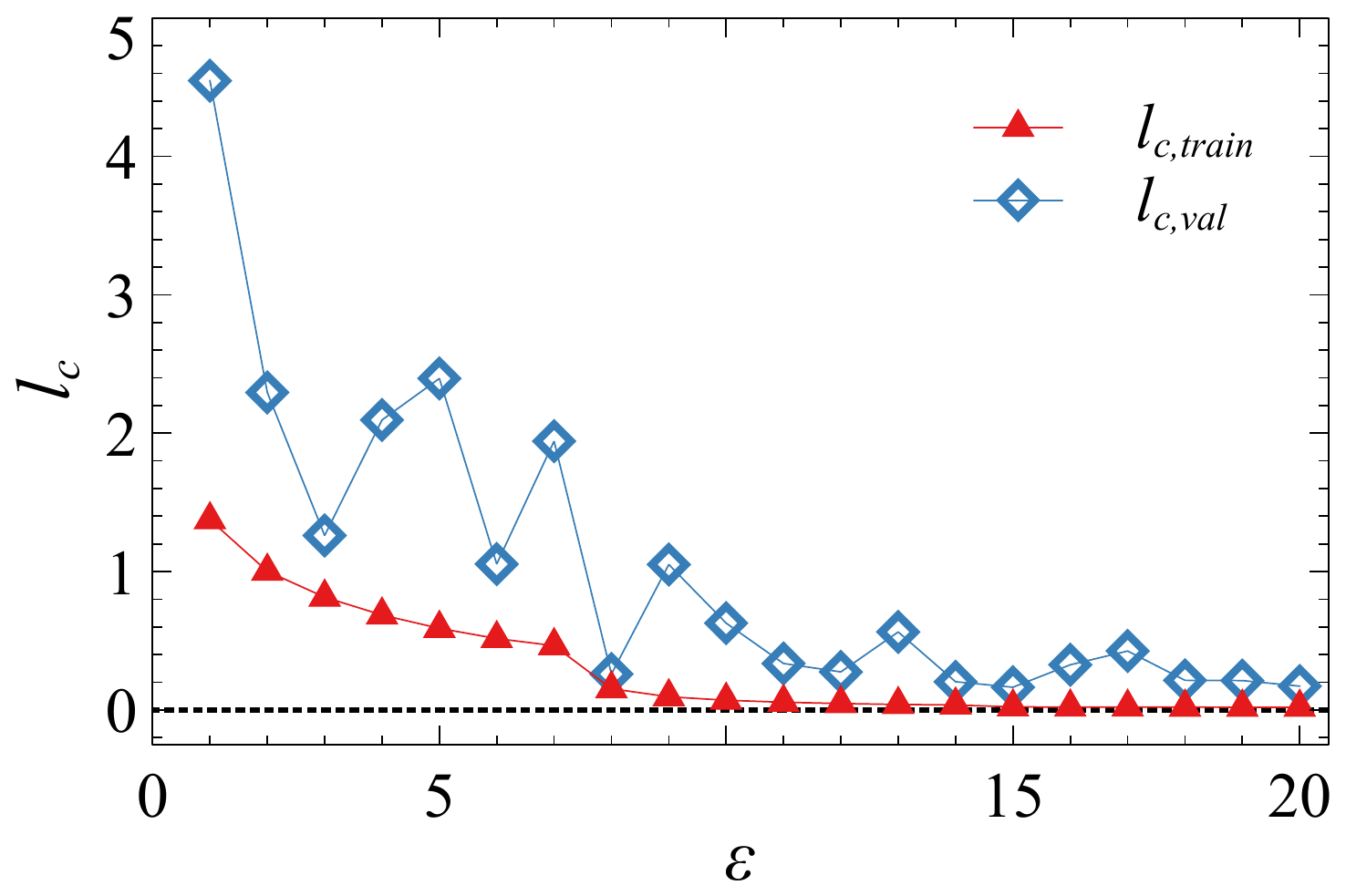}\hspace{+3ex}
    \raisebox{0.30\columnwidth}{(d)}\hspace*{-1ex}
    \includegraphics[width=0.35\columnwidth]{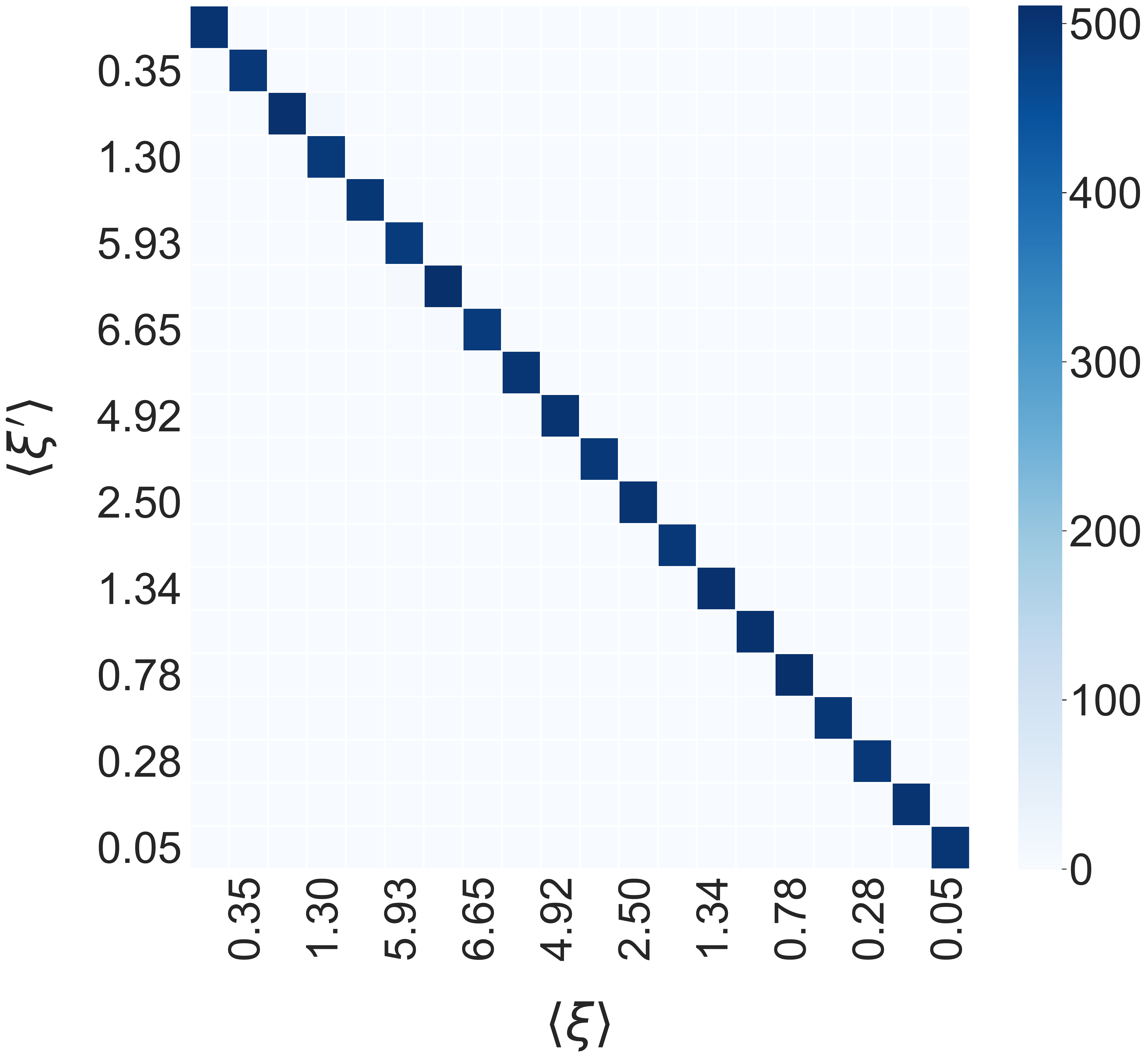}\\
    \raisebox{0.30\columnwidth}{(e)}\hspace*{+1ex}
    \includegraphics[width=0.48\columnwidth]{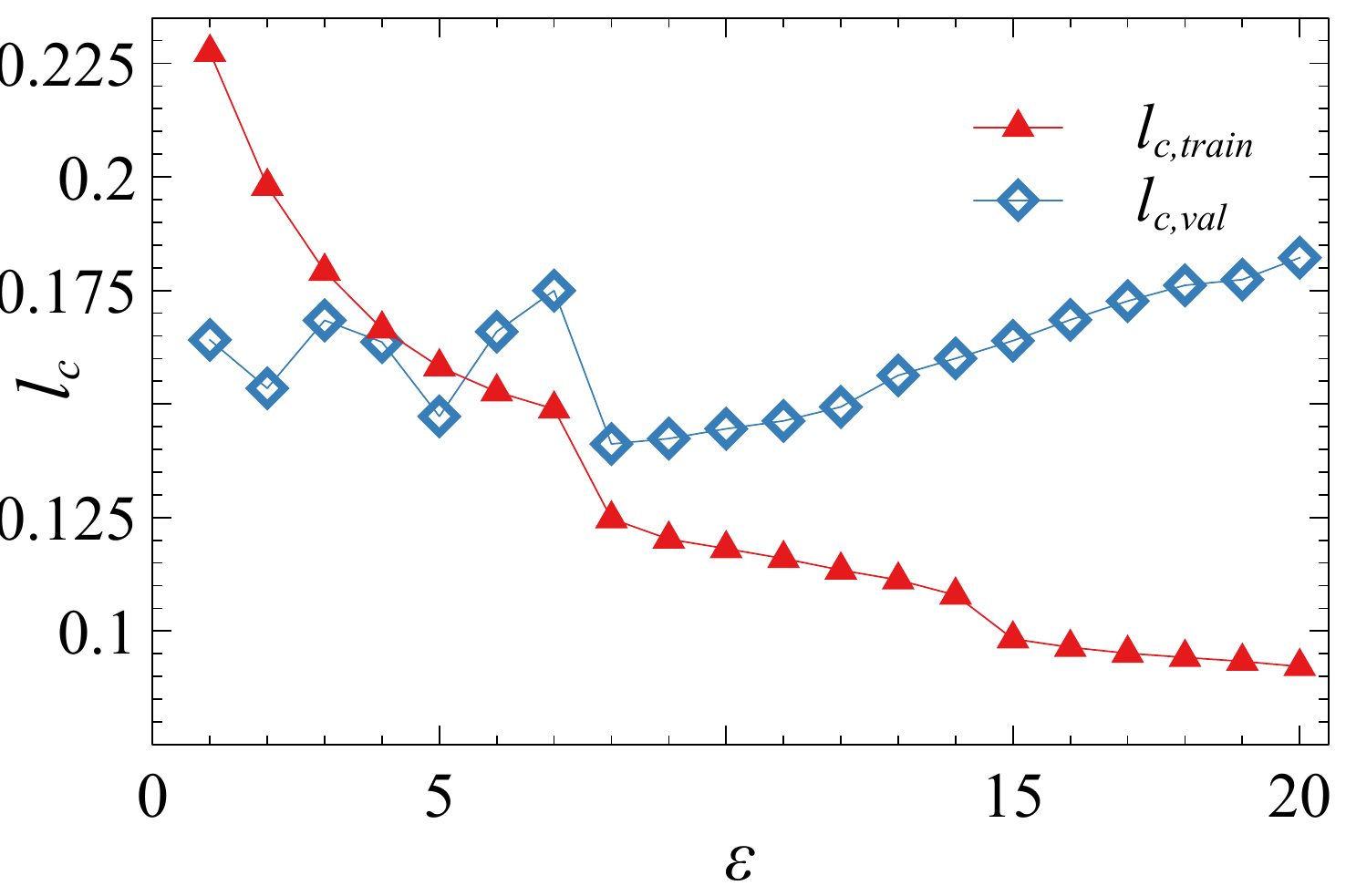}\hspace{+3ex}
    \raisebox{0.30\columnwidth}{(f)}\hspace*{-1ex}
    \includegraphics[width=0.35\columnwidth]{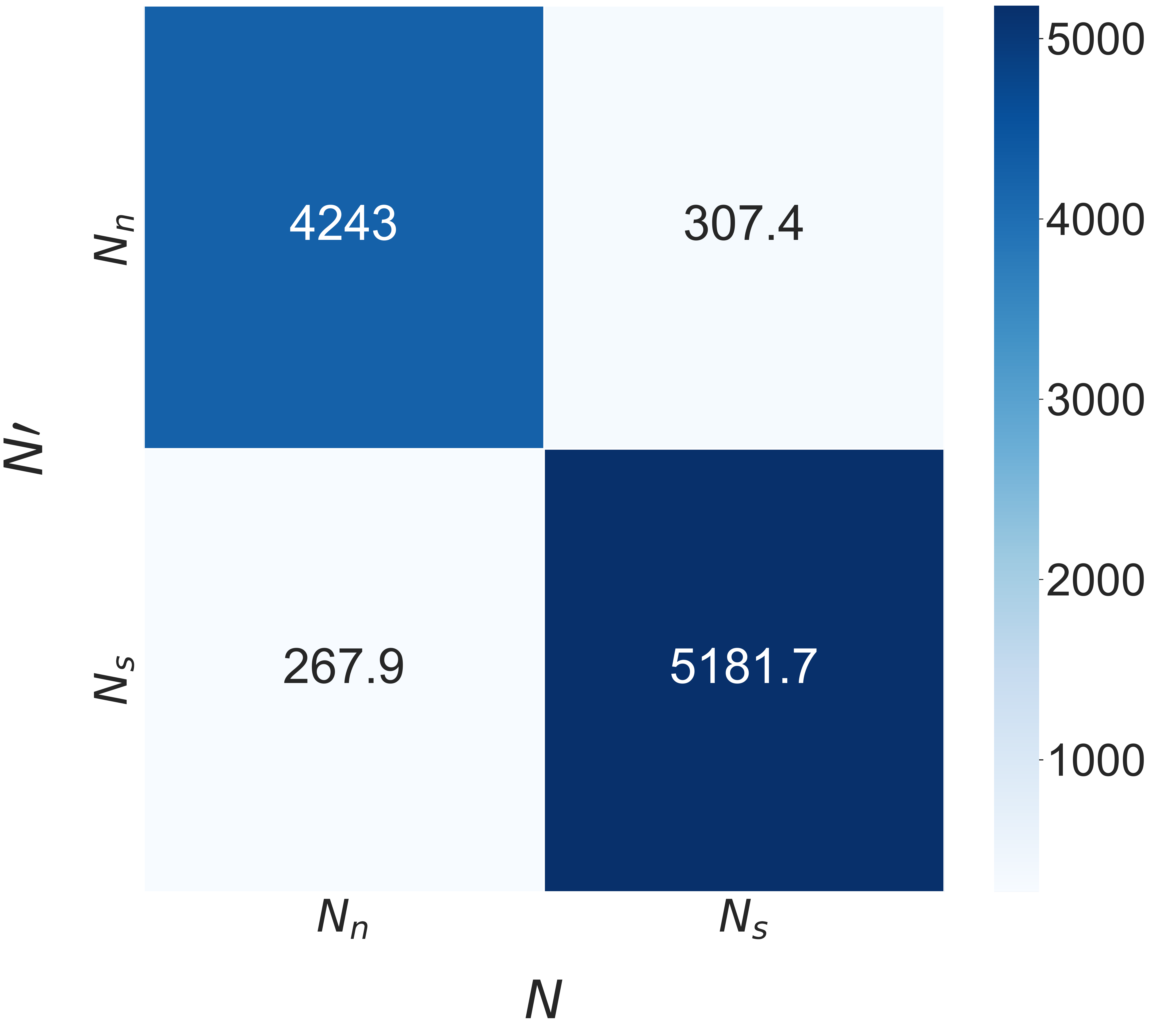}
%\vspace*{-1.5mm}
    \caption{(a,c,e) Dependence of losses $l_\mathrm{c,train}$ and $l_\mathrm{c,val}$ on the number of epochs $\epsilon$ for classification according to (a) $p$'s, (c) $\xi$'s and (e) spanning/non-spanning. The diamonds (blue open) denote $l_\mathrm{c,train}$ while the triangles (red solid) show $l_\mathrm{c,val}$. The thin lines are guides to the eye only. The horizontal dotted lines in (a,c) indicate $l_\mathrm{c}=0.$
    \revision{(b,d,f)} Confusion matrices for the validation data used in (a,c,e) at $\epsilon=20$. The true labels for $p$, $\xi$ and $N_s$/$N_n$ are indicated on the horizontal axes while the predicted labels are given on the vertical axes. The colour scale represents the number of samples in each matrix entry, for (f) these are given explicitly.
    }
    \label{fig:ml-class-density-learningcurve}
    \label{fig:ml-class-corrlen-learningcurve}
    \label{fig:ml-class-spanning-learningcurve}
\vspace*{-2.5mm}
\end{figure*}

Figure \ref{fig:ml-class-density-learningcurve}(a) shows the training and validation losses for classifying densities.
%achieved for the pre-trained {\sc ResNet} described in section \ref{sec:modelmethods} when applied to the $20$ densities $p= 0.1$, $0.2$, $\ldots$, $0.5$, $0.55$, $0.56$, $\ldots$, $0.66$, $0.7$, $\ldots 0.9$.
%%%%%%%%%%%%%%%%%%%%%%%%%%%%%%%%%%%%%%%%%%%%%%%%%%%%%%%%%%%%%%%%%%%%%%%%%%%%%%%%
%%%%%%%%%%%%%%%%%%%%%%%%%%%%%%%%%%%%%%%%%%%%%%%%%%%%%%%%%%%%%%%%%%%%%%%%%%%%%%%%
%
% %%%%%%%%%%%%%%%%%%%%%%%%%%%%%%%%%%%%%%%%%%%%%%%%%%%%%%%%%%%%%%%%%%%%%%%%%%%%%%%%
% \begin{figure*}[tb]
%     \centering%
%     (a)\includegraphics[width=0.5\columnwidth]{Classification/class_dens_loss.pdf}
%     (b)\includegraphics[width=0.42\columnwidth]{Classification/cm_density_class.pdf}
%     \caption{
%     (a) Dependence of losses $l_\mathrm{c,train}$ and $l_\mathrm{c,val}$ on the number of epoch $\epsilon$ for classification according to densities. The diamonds (blue open) denote $l_\mathrm{c,train}$ while the triangles (red solid) show $l_\mathrm{c,val}$. The thin lines are guides to the eye only. The horizontal dotted line indicates $l_\mathrm{c}=0.$
%     (b) Confusion matrix for the validation data used in (a) at $\epsilon=20$. The true labels $p=0.5, \ldots, 0.9$ are indicated on the horizontal axis while the predicted labels $p'$ are given on the vertical axis. The color scale represents the number of samples in each $(p,p')$ entry.
%     }
%     \label{fig:ml-class-density-learningcurve}
% \end{figure*}
% %%%%%%%%%%%%%%%%%%%%%%%%%%%%%%%%%%%%%%%%%%%%%%%%%%%%%%%%%%%%%%%%%%%%%%%%%%%%%%%%
%
An excellent training result is corroborated by a perfectly diagonal confusion matrix shown in Fig.\ \ref{fig:ml-class-density-learningcurve}(b). We note that this also holds when the spacing of densities is reduced as for the values $p= 0.55, 0.56, \ldots, 0.66$ around $p_c$. 
%
%\subsubsection{Correlation lengths}
%
We now repeat the procedure of classification, but replace the density $p$ labels with the computed average correlation lengths, $\langle\xi(p)\rangle$. These are $\langle\xi(0.1)\rangle, \langle\xi(0.2)\rangle, \ldots, \langle\xi(0.9)\rangle$ corresponding to 
$0.18\pm 0.07$, $0.35\pm 0.07$, $0.65\pm 0.05$, $1.30\pm 0.05$, $3.36\pm 0.07$, $5.93\pm 0.06$, $6.48\pm 0.04$, $6.65\pm 0.07$, $6.18\pm 0.05$, $4.92\pm 0.04$, $3.56\pm 0.06$, $2.50\pm 0.03$, $1.78\pm 0.05$, $1.34\pm 0.07$, $1.01\pm 0.07$, $0.78\pm 0.05$, $0.62\pm 0.04$, $0.28\pm 0.04$, $0.08\pm 0.07$, $0.05\pm 0.05$, respectively, in Fig.\ \ref{fig:schematics}(d).
In Figs.\ \ref{fig:ml-class-corrlen-learningcurve}(c, d) we show $l_\mathrm{c,train}(\epsilon)$, $l_\mathrm{c,val}(\epsilon)$ and the confusion matrix, respectively.
% %%%%%%%%%%%%%%%%%%%%%%%%%%%%%%%%%%%%%%%%%%%%%%%%%%%%%%%%%%%%%%%%%%%%%%%%%%%%%%%%%
% \begin{figure*}[tb]
%     \centering%
%     (a)\includegraphics[width=0.5\columnwidth]{Classification/plot_loss_class_corr_l.pdf}
%     (b)\includegraphics[width=0.42\columnwidth]{Classification/avg_cm_10.pdf}
%     \caption{
%     (a) Dependence of $l_\mathrm{c,train}$ and $l_\mathrm{c,val}$ on $\epsilon$ for classification by $\langle\xi\rangle$. Diamonds, triangles and lines are as in Fig.\ \ref{fig:ml-class-density-learningcurve}.
%     (b) Confusion matrix for the validation data at $\epsilon=20$. The true values $\xi_1, \ldots, \xi_{20}$ as in Fig.\ \ref{fig:schematics}(d) are indicated on the horizontal axis while the predicted values $\xi_{i}'$ are on the vertical axis. The color scale denoted the number of samples as in Fig.\ \ref{fig:ml-class-density-learningcurve}.
%     }
%     \label{fig:ml-class-corrlen-learningcurve}
% \end{figure*}
% %%%%%%%%%%%%%%%%%%%%%%%%%%%%%%%%%%%%%%%%%%%%%%%%%%%%%%%%%%%%%%%%%%%%%%%%%%%%%%%%%
We again use the full set of $20$ classes for the classification. As we can see in Fig.\ \ref{fig:ml-class-corrlen-learningcurve}(c), the loss for the classification is as good as for the density classification shown in Fig.\ \ref{fig:ml-class-density-learningcurve}(a). Of course this should not be surprising since we effectively just replaced the $p$ labels with $\xi$ labels.

%%%%%%%%%%%%%%%%%%%%%%%%%%%%%%%%%%%%%%%%%%%%%%%%%%%%%%%%%%%%%%%%%%%%%%%%%%%%%%%%%
%\subsubsection{Spanning non-spanning}

The percolation transition distinguishes non-spanning clusters at $p < p_c$ from spanning clusters at $p \geq p_c$. We therefore now also train the {\sc ResNet} with all percolation  data, but just using their classification in non-spanning and spanning percolation clusters as  labels. In Figs.\ \ref{fig:ml-class-spanning-learningcurve}(e) and (f) we show $l_\mathrm{c,train}(\epsilon)$, $l_\mathrm{c,val}(\epsilon)$ and the confusion matrix, respectively. 
% %%%%%%%%%%%%%%%%%%%%%%%%%%%%%%%%%%%%%%%%%%%%%%%%%%%%%%%%%%%%%%%%%%%%%%%%%%%%%%%%%
% \begin{figure*}[tb]
%     \centering
%     (a)\includegraphics[width=0.5\columnwidth]{Classification/loss_class_span.pdf}
%     (b)\includegraphics[width=0.42\columnwidth]{Classification/cm_span_class.pdf}
%     % \caption{(a) Evolution of the accuracy during the training of the 2 classes 
%     % (b) Confusion matrix obtained after training of the pretrained ResNet18 for 2 classes (spanning or non spanning)}
%     \caption{
%     (a) Dependence of $l_\mathrm{c,train}$ and $l_\mathrm{c,val}$ on $\epsilon$ for classification by spanning/non-spanning. Diamonds, triangles and lines are as in Figs.\ \ref{fig:ml-class-density-learningcurve} and \ref{fig:ml-class-corrlen-learningcurve}.
%     (b) Confusion matrix for the validation data at $\epsilon=7$. The true labels $N_s$, for spanning, and $N_n$, for non-spanning, are on the horizontal axis while the predicted labels $N_s'$ and $N_s'$ are given on the vertical axis. The color scale and the numbers in each entry of the confusion matrix denote the number of spanning and non-spanning clusters. }
%     \label{fig:ml-class-spanning-learningcurve}
% \end{figure*}
% %%%%%%%%%%%%%%%%%%%%%%%%%%%%%%%%%%%%%%%%%%%%%%%%%%%%%%%%%%%%%%%%%%%%%%%%%%%%%%%%%
The confusion matrix shows good recognition for the majority of cases. Nevertheless, we also see that about $5 \%$ of the samples get misclassified, i.e.,  $298$ of $4565$ non-spanning clusters are classified as spanning while $279$ of $5434$ spanning clusters are wrongly identified as non-spanning. This suggests that the classification routine is struggling with correctly identifying the spanning cluster.

%%%%%%%%%%%%%%%%%%%%%%%%%%%%%%%%%%%%%%%%%%%%%%%%%%%%%%%%%%%%%%%%%%%%%%%%%%%%%%%%%
%\subsection{Regression}

%%%%%%%%%%%%%%%%%%%%%%%%%%%%%%%%%%%%%%%%%%%%%%%%%%%%%%%%%%%%%%%%%%%%%%%%%%%%%%%%%
%\subsubsection{Density}

\begin{figure*}[htb]
    \centering%
    \raisebox{0.27\columnwidth}{(a)}\hspace*{-2ex}
    % \hspace*{-0.7cm}
    \includegraphics[width=0.44\columnwidth]{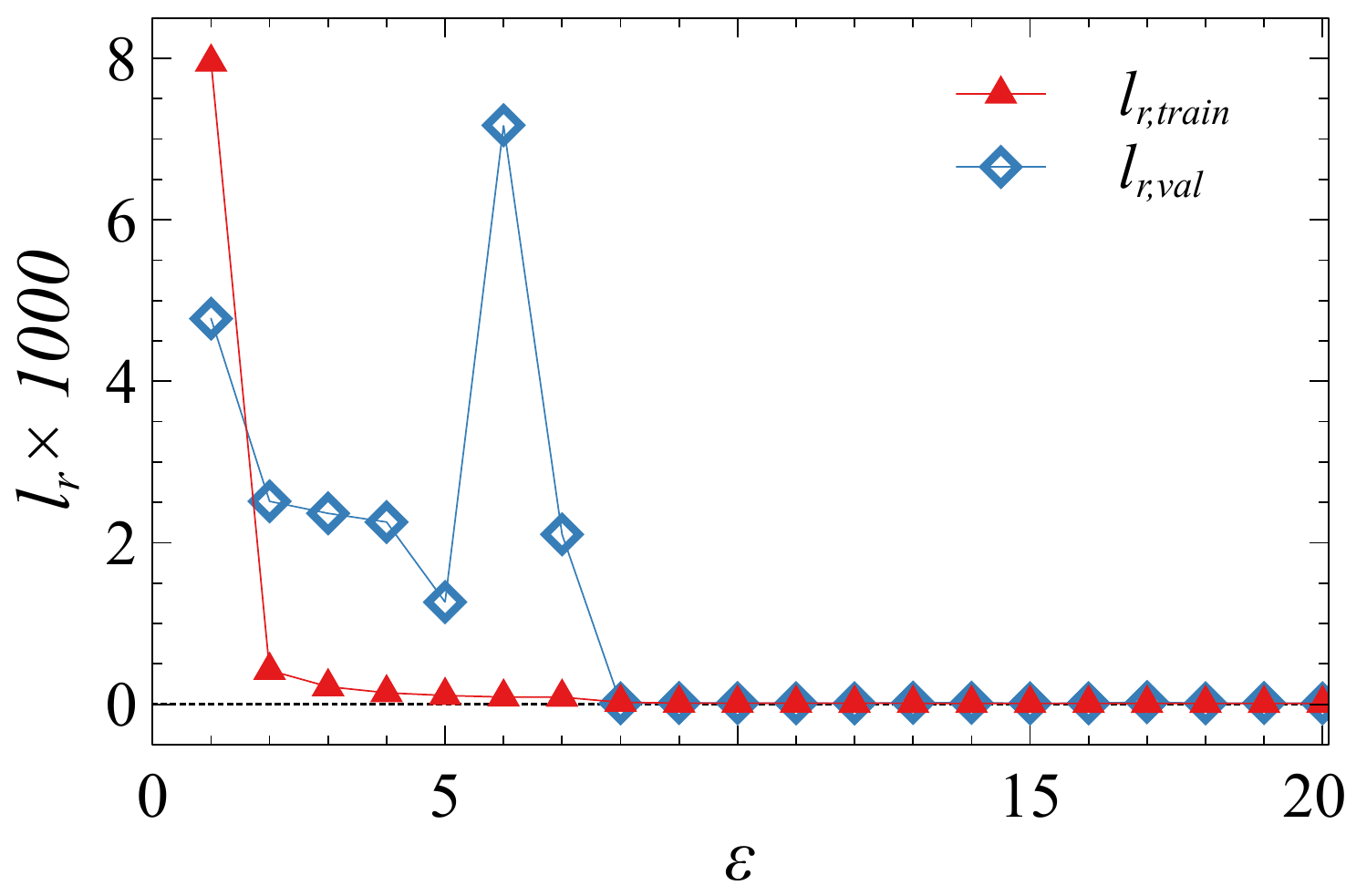}
    \hspace*{+3ex}
    \raisebox{0.27\columnwidth}{(b)}\hspace*{-1ex}
    \includegraphics[width=0.44\columnwidth]{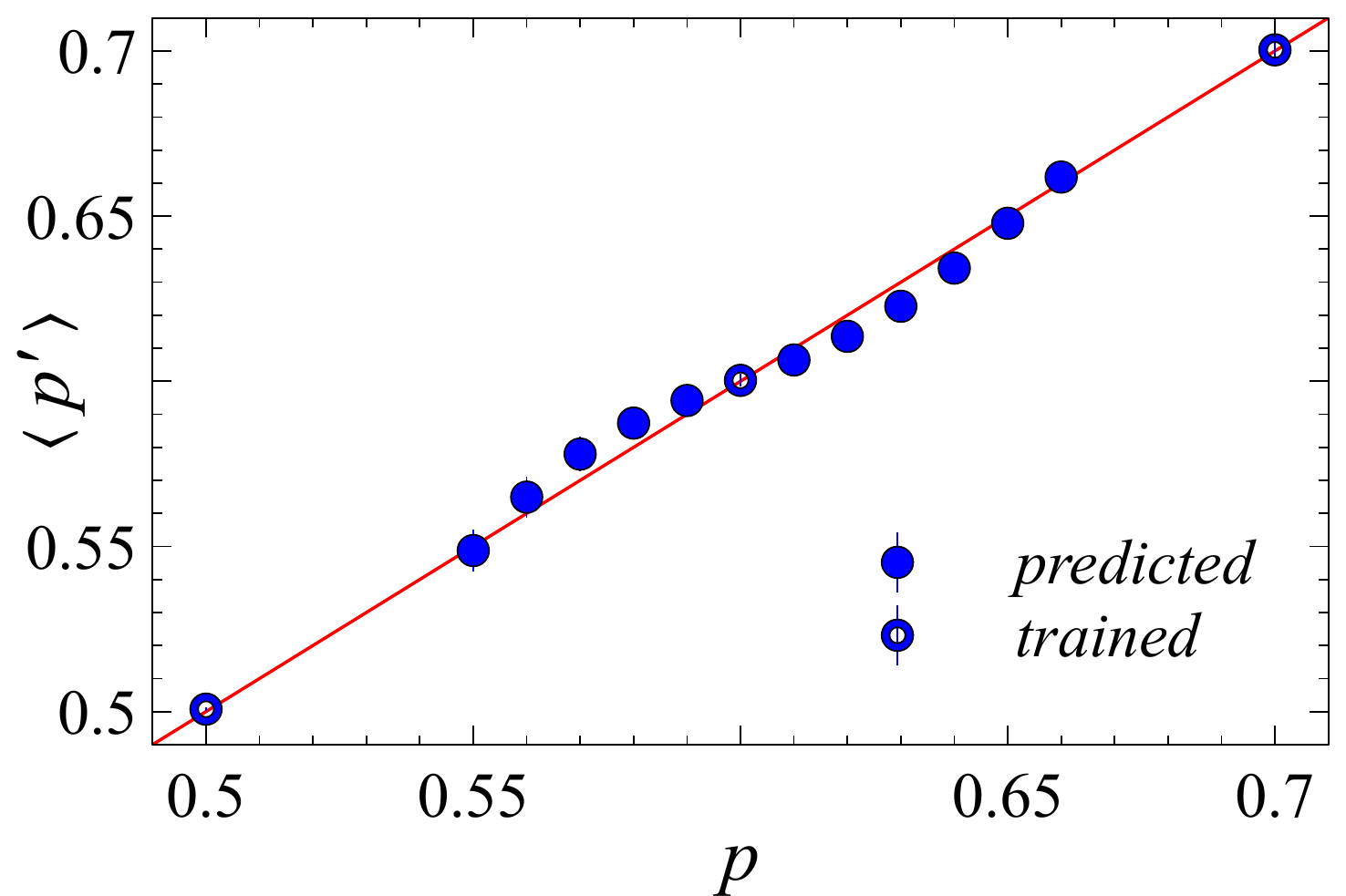}\\
    \raisebox{0.27\columnwidth}{(c)}\hspace*{-2ex}
    % \hspace*{-0.7cm}
    \includegraphics[width=0.44\columnwidth]{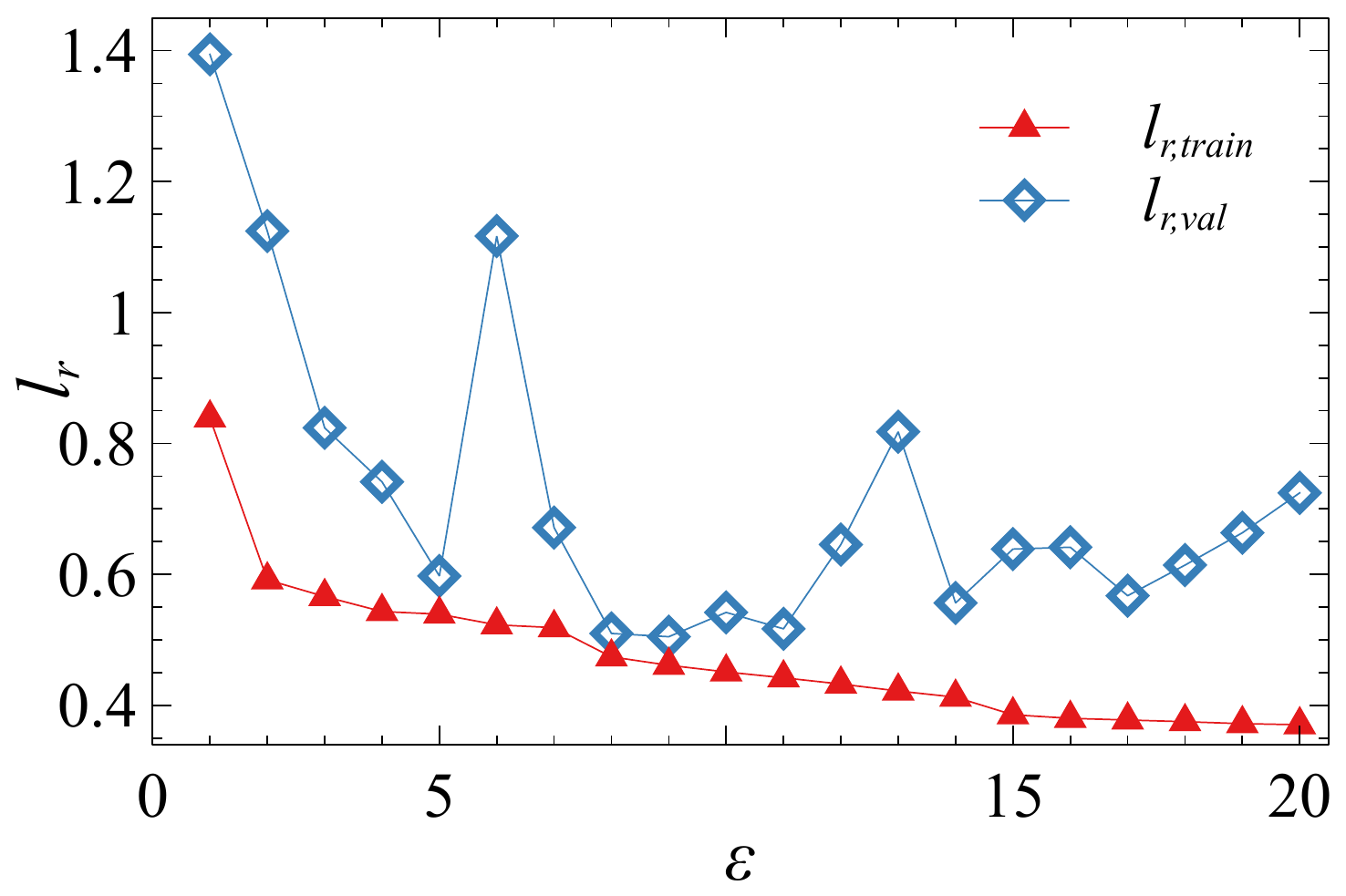}
    \hspace*{+3ex}
    \raisebox{0.27\columnwidth}{(d)}\hspace*{-1ex}
    \includegraphics[width=0.44\columnwidth]{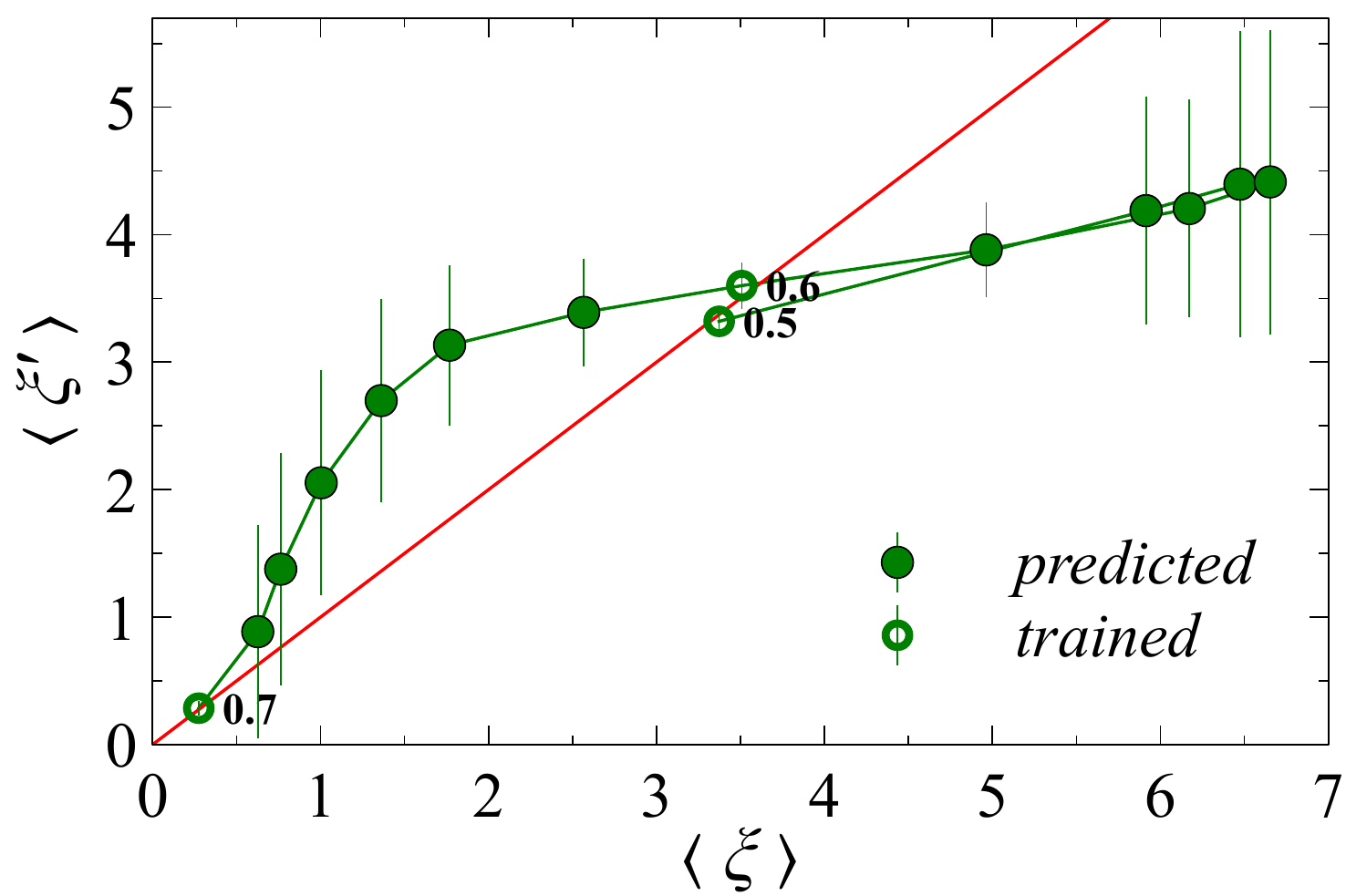}
%\vspace*{-1.5mm}
    \caption{(a,c) Dependence of $l_\mathrm{r,train}$ and $l_\mathrm{r,val}$ on $\epsilon$ for the regression task with (a) $p$'s and (c) $\xi$'s. Similar to Fig.\ \ref{fig:ml-class-density-learningcurve}, triangles denote $l_\mathrm{r,train}$ and diamonds show $l_\mathrm{r,val}$. Lines are as before.
    (b, d) Dependence of the true (b) $p$ on $p'$ and (d) $\xi$ on $\xi$'. The red lines denotes the perfect predictions $p=p'$ and $\langle\xi\rangle=\langle\xi\rangle'$. The three open circles highlight (c) $p=0.5$, $0.6$ and $0.7$ and, respectively, (d) $\langle\xi\rangle=3.36\pm 0.07$, $3.56 \pm 0.06$ and $0.28 \pm 0.04$ included in the training sets while the solid circles are the results of the regression.
    }
\vspace*{-2.5mm}
    \label{fig:ml-reg-dens-losscurve}
    \label{fig:ml-reg-corrlen-losscurve}
\end{figure*}

The classification approach is in principle restricted in its predictive accuracy for $p$ by the number of available classes. While one can of course try to increase the number of classes, this will also lead to an increased chance in misclassifying percolation states from close by classes. The regression method outlined in section \ref{sec:modelmethods} allows to predict the $p$ value for a given percolation state directly. We again start from a pre-trained {\sc ResNet} in Fig.\ \ref{fig:ml-reg-dens-losscurve}(a) we show the training and validation losses, $l_\mathrm{r,train}$ and $l_\mathrm{r,val}$, respectively.
%%%%%%%%%%%%%%%%%%%%%%%%%%%%%%%%%%%%%%%%%%%%%%%%%%%%%%%%%%%%%%%%%%%%%
%%%%%%%%%%%%%%%%%%%%%%%%%%%%%%%%%%%%%%%%%%%%%%%%%%%%%%%%%%%%%%%%%%%%%%%
% %%%%%%%%%%%%%%%%%%%%%%%%%%%%%%%%%%%%%%%%%%%%%%%%%%%%%%%%%%%%%%%%%%%%%
% \begin{figure*}[tb]
%     \centering%
%     (a)\includegraphics[width=0.45\columnwidth]{Regression/plot_loss_dens_reg.pdf}
%     (b)\includegraphics[width=0.45\columnwidth]{Regression/pred_reg_dens.pdf}
%     \caption{
%     (a) Dependence of $l_\mathrm{r,train}$ and $l_\mathrm{r,val}$ on $\epsilon$ for the regression task with densities. Similar to Fig.\ \ref{fig:ml-class-density-learningcurve}, triangles denote $l_\mathrm{r,train}$ and diamonds show $l_\mathrm{r,val}$. Lines are as before.
%     (b) Dependence of the true labels $p$ on $p'$. The thin line denotes a perfect prediction $p=p'$. The three open circles highlight $p=0.5$, $0.6$ and $0.7$ included in the training set while the solid circles are the results of the regression.
%     }
%     \label{fig:ml-reg-dens-losscurve}
% \end{figure*}
% %%%%%%%%%%%%%%%%%%%%%%%%%%%%%%%%%%%%%%%%%%%%%%%%%%%%%%%%%%%%%%%%%%%%%%%
We train the {\sc ResNet} for the evenly-spaced $9$ densities $p= 0.1, 0.2, \ldots, 0.9$. We then use the data from the $12$ densities $p= 0.55, 0.56, \ldots, 0.66$ as test of the regression model. We note that in this approach, the $p=0.6$ data is used both in the training/validation cycle as well as in the test.
In Fig.\ \ref{fig:ml-reg-dens-losscurve}(b) we show the dependence of the true density $p$ as a function of the average predicted density $p'$. 
We see that the predicted $p'$ values closely follow the true $p$'s. However, we also find that the predictions cluster somewhat around $p=0.6$, a $p$ value used in the training.

%%%%%%%%%%%%%%%%%%%%%%%%%%%%%%%%%%%%%%%%%%%%%%%%%%%%%%%%%%%%%%%%%%%%%%%%%%%%%%%%%
%\subsubsection{Correlation lengths}

We can also perform a regression study using the correlation lengths $\xi$ as regression target. However, we now have a distribution of computed $\xi$ values for each $p$ whereas for $p$-regression, each sample has exactly the correct $p$ value. For a given $p$ value, we therefore construct an average value $\langle \xi \rangle$. In order to capture this behaviour correctly, we train our {\sc ResNet} on the individual $\xi$ values for each sample at given $p$. Similarly, when computing the predicted $\xi'$ values, we average the individual predictions for given $p$ to construct $\langle \xi'\rangle$. 
As before for $p$ regression, we train using the $\xi$ values of the percolation states for densities $p=0.1, 0.2,\ldots,0.9$ and predict $\xi'$ from the $12$ densities $p=0.55,0.56,\ldots,0.66$. While all the previous training were performed on percolation states where empty sites were denoted by 0 and occupied site by 1, this dataset did not provide exploitable results. We therefore decided to train with percolation states containing the numbered cluster found through the Hoshen-Kopelman algorithm \cite{PhysRevB.14.3438}.
%
%In Fig.\ \ref{fig:ml-reg-corrlen-losscurve}(c) we show the loss functions $l_\mathrm{train}$ and $l_\mathrm{val}$. 
We obtain the dependence of the average predicted label $\langle \xi' \rangle$ on the average true label $\langle \xi\rangle$ as shown in Fig.\ \ref{fig:ml-reg-corrlen-losscurve}(d).
% %%%%%%%%%%%%%%%%%%%%%%%%%%%%%%%%%%%%%%%%%%%%%%%%%%%%%%%%%%%%%%%%%%%%%%%%%%%%%%%%%
% \begin{figure*}[tb]
%     \centering
%     (a)\includegraphics[width=0.45\columnwidth]{Regression/plot_loss_regres_corr.pdf}
%     (b)\includegraphics[width=0.45\columnwidth]{Regression/meantruevspredicted_corr.jpg}
%     \caption{(a) Dependence of $l_\mathrm{r,train}$ and $l_\mathrm{r,val}$ on $\epsilon$ for the $\xi$ regression. Similarly to Fig.\ \ref{fig:ml-reg-dens-losscurve} triangles give $l_\mathrm{r,train}$, diamonds represent $l_\mathrm{r,val}$ and lines are as before guides to the eye.
%     (b) The thin lines gives $\langle \xi\rangle=\langle \xi'\rangle$ .
%     }
%     \label{fig:ml-reg-corrlen-losscurve}
% \end{figure*}
% %%%%%%%%%%%%%%%%%%%%%%%%%%%%%%%%%%%%%%%%%%%%%%%%%%%%%%%%%%%%%%%%%%%%%%%%%%%%%%%%%
In Fig.\  \ref{fig:ml-reg-corrlen-losscurve}(b), we see that there are important deviations from a perfect regression curve $\langle \xi' \rangle= \langle \xi\ \rangle$ with $\langle \xi \rangle$ corresponding to trained configuration $p=0.5$, $p=0.6$ and $p=0.7$ on the line of perfect regression. 
%We observe that training with a dataset containing information on the cluster does not help the network to successfully predict correlation length. 

%%%%%%%%%%%%%%%%%%%%%%%%%%%%%%%%%%%%%%%%%%%%%%%%%%%%%%%%%%%%%%%%%%%%%%%%%%%%%%%%%
\section{Conclusions}
\label{sec:concl}
%%%%%%%%%%%%%%%%%%%%%%%%%%%%%%%%%%%%%%%%%%%%%%%%%%%%%%%%%%%%%%%%%%%%%%%%%%%%%%%%%
% p + xi classification work
We have shown that CNNs, and in particular pretrained {\sc ResNets}, are able to successfully classify percolation states according to densities $p$ and correlation lengths $\xi$ in the case of the 2D site percolation on a square lattice. 
% spanning-non-spanning has problems
When trying to classify spanning versus non-spanning percolation states, the results are less convincing. While the majority of samples are being correctly classified, there are about $5\%$ of percolation states which are wrongly classified. This suggests that the {\sc ResNet} is not able to identify the percolating/non-percolating nature of clusters correctly.
% p + xi regression works
For the regression analysis, when trained with evenly spaced densities $p=0.1$, $0.2$, $\ldots$, $0.9$, we find good quantitative predictions, even for much closer spaced densities $p= 0.55$, $0.56$, $\ldots$, $0.66$. For the correlation length, the classification gave us perfect classes predictions, but the regression analysis showed that the network might not understand and therefore predict correctly correlation lengths. 

% does ML recognise spanning clusters or does it count occupation (density)? 
In order to make sense of these results, one might want to ask whether the ML image recognition tools employed here simply count the number of occupied and unoccupied sites while disregarding the existing of a percolating cluster. Clearly, this would be consistent with the precision of the classification and regression results for $p$ and $\xi$. However, then we would not expect, e.g., the deviations from $p=p'$ and $\xi=\xi'$ seen in Fig.\ \ref{fig:ml-reg-corrlen-losscurve}(b,d). Similarly, when looking at which clusters are wrongly classified during the spanning/non-spanning classification, we find that these are mostly \revision{spanning for $p\lesssim p_c$ and non-spanning for $p\gtrsim p_c$}. Clearly, more studies are needed to clarify these issues.

% Looking at the spanning/non-spanning classification we see an overall satisfying result with the . However the misclassified percolation state might gave us an hint on a possible lead explaining the disappointing result obtained with the regression training on the correlation length. The notion of connectivity might not be one that the network recognise, the main feature that the network seem to pick up is the density. 

% % VAE
% A possible way to corroborate this lead would be to use unsupervised learning techniques, in particular generative method such as the Variational Autoencoder (VAE).

%%%%%%%%%%%%%%%%%%%%%%%%%%%%%%%%%%%%%%%%%%%%%%%%%%%%%%%%%%%%%%%%%%%%%%%%%%%%%%%%%
\section*{References}
%\bibliographystyle{iopart-num}
%\bibliography{ML-refs-DB_corrected.bib}

\providecommand{\newblock}{}

\end{document}